\documentclass[aps, prd, preprint, amsfonts,
  amssymb, amsmath, showpacs, a4paper,nofootinbib,superscriptaddress]{revtex4-1}
  
\usepackage[T1]{fontenc} 

\usepackage{slashed}
\usepackage{braket}

\newcommand{\be}{\begin{equation}}
\newcommand{\ee}{\end{equation}}
\newcommand{\bea}{\begin{eqnarray}}
\newcommand{\eea}{\end{eqnarray}}

\usepackage{color}

\begin{document}

\title{Spontaneous symmetry breaking and the Goldstone theorem in\\ non-Hermitian field theories}

\author{Jean Alexandre}
\email{jean.alexandre@kcl.ac.uk}
\affiliation{Department of Physics, King's College London,\\ 
London WC2R 2LS, United Kingdom}

\author{John Ellis}
\email{john.ellis@cern.ch}
\affiliation{Department of Physics, King's College London,\\ 
London WC2R 2LS, United Kingdom}
\affiliation{National Institute of Chemical Physics \& Biophysics, R\"avala 10, 10143 Tallinn, Estonia}
\affiliation{Theoretical Physics Department, CERN, CH-1211 Geneva 23, Switzerland}

\author{Peter Millington}
\email{p.millington@nottingham.ac.uk}
\affiliation{School of Physics and Astronomy, University of Nottingham,\\ Nottingham NG7 2RD, United Kingdom\vspace{1em}}

\author{Dries Seynaeve}
\email{dries.seynaeve@kcl.ac.uk}
\affiliation{Department of Physics, King's College London,\\ 
London WC2R 2LS, United Kingdom}

\begin{abstract}

We demonstrate the extension to $\mathcal{PT}$-symmetric field theories of the Goldstone theorem, confirming that
the spontaneous appearance of a field vacuum expectation value via minimisation of the effective potential
in a non-Hermitian model is accompanied by a massless scalar boson. 
Laying a basis for our analysis, we first show how the conventional quantisation of the path-integral formulation
of quantum field theory can be extended consistently to a non-Hermitian model by
considering $\mathcal{PT}$ conjugation instead of Hermitian conjugation. The extension of the Goldstone theorem
to a $\mathcal{PT}$-symmetric field theory is made possible by the existence of a conserved current that does not, however, 
correspond to a symmetry of the non-Hermitian Lagrangian. In addition to extending the 
proof of the Goldstone theorem to a $\mathcal{PT}$-symmetric theory, we exhibit a specific example in which
we verify the existence of a massless boson at the tree and one-loop levels.\\
~~\\
KCL-PH-TH/2018-18, CERN-PH-TH/2018-117
~~\\
August 2018

\end{abstract}

\maketitle


\section{Introduction}

The conventional formulations of quantum mechanics and quantum field theory (QFT) have generally been
based on Hermitian Hamiltonians and Lagrangians, respectively. However, in recent years it has been
established that one can also consistently formulate non-Hermitian, $\mathcal{PT}$-symmetric
quantum-mechanical models~\cite{Bender:2005tb}, and the possibility of a smooth transition between 
Hermitian and $\mathcal{PT}$-symmetric phases in quantum mechanics is described in Ref.~\cite{Znojil:2017viu}.
Non-Hermitian QFTs have also been studied in various contexts. For example, a model with an $i\phi^3$ scalar interaction
was studied in Refs.~\cite{Blencowe:1997sy,Jones:2004gp,Bender:2013qp,Shalaby:2017wux},
and it was shown that a meaningful unbounded effective potential can be obtained in the framework of 
$\mathcal{PT}$-symmetric QFT~\cite{Bender:2015uxa}.

A $\mathcal{PT}$-symmetric QFT involving a non-Hermitian fermion mass term 
$\mu\bar\psi\gamma^5\psi$ was introduced in Ref.~\cite{Bender:2005hf}.
This model was studied further in Ref.~\cite{Alexandre:2015oha}, where the existence of a conserved current was
demonstrated and shown to ensure the consistency of $\mathcal{PT}$ symmetry with unitarity. 
This non-Hermitian mass term has been used for alternative descriptions of neutrino masses~\cite{JonesSmith:2009wy,Alexandre:2015kra} (see also Ref.~\cite{Alexandre:2017fpq} for a summary) or dark matter \cite{Rodionov:2017dqt}. Non-Hermitian extensions of conventional QFT have also been applied to 
neutrino oscillations \cite{Ohlsson:2015xsa} and to decays of the Higgs boson \cite{Korchin:2016rsf}.
Interesting studies have been done in Ref.~\cite{Chernodub:2017lmx} of a non-Hermitian fermionic model on the lattice,
which allows for a different number of left-handed and right-handed excitations, consistent with the fermionic 
current density derived in Ref.~\cite{Alexandre:2015oha}. We also note that the confinement phase transition in 
QCD has been related to $\mathcal{PT}$-symmetry properties of ghost fields in Ref.~\cite{Raval:2018kqg}.

An intriguing feature of Ref.~\cite{Alexandre:2015oha} was the discovery that the existence of a conserved
current in a $\mathcal{PT}$-symmetric QFT does not correspond to a symmetry of the Lagrangian ${\cal L}$, 
but rather to a specific transformation of ${\cal L}$ that is related to the non-Hermitian part of the action.
$\mathcal{PT}$-symmetric QFTs evade Noether's theorem~\cite{Noether} in the sense that symmetries of the 
Lagrangian do not give rise to conserved currents. Revisiting Noether's derivation, one finds that there exist 
conserved currents for non-Hermitiam theories, but these correspond to transformations that must effect a 
particular non-trivial variation of the Lagrangian, which vanishes only in the Hermitian limit.
This observation raises the interesting question 
of whether there is an analogue in $\mathcal{PT}$-symmetric QFT of spontaneous symmetry breaking and,
if so, whether the breaking of a global symmetry is accompanied by a massless Goldstone mode, as in Hermitian QFT~\cite{GSW1,GSW2,GSW3}. The answers
provided in this paper are that the existence of a massless Goldstone mode can be shown from current 
conservation and does not require the Lagrangian to be invariant under the corresponding
field transformation. Nevertheless, there is a symmetry of the Lagrangian, which 
is spontaneously broken by the choice of a specific vacuum.

However, before addressing these questions, we first discuss some basic issues in the formulation of a
non-Hermitian QFT, which require a consistent procedure for quantisation of the path integral. This
is based on the existence of a complete set of real energy eigenstates, which allow the introduction of a
saddle point about which the integration of quantum fluctuations is well defined. To this end, we
show how this conventional quantisation of the path integral can be extended consistently to a 
non-Hermitian scalar QFT by considering $\mathcal{PT}$ conjugation instead of Hermitian conjugation.
We perform the calculation of the one-loop effective action explicitly for a generic case and, 
assuming that the only source of non-Hermiticity is a mass term, we show that the theory is asymptotically Hermitian.

We then prove an extension of the Goldstone theorem for this non-Hermitian QFT, showing that
the spontaneous appearance of a field vacuum expectation value via minimisation of the effective potential
is accompanied by the appearance of a massless scalar mode, whose
existence is linked to the presence of a conserved current in this $\mathcal{PT}$-symmetric QFT.
We confirm the existence of the massless Goldstone mode by explicit calculations at both the tree and one-loop levels.

The layout of this paper is as follows. In Sec.~\ref{sec:model}, we review the variational procedure 
(originally described in Ref.~\cite{Alexandre:2017foi} and summarized in Ref.~\cite{Alexandre:2017erl}) for the complex scalar model that forms the focus of this work. 
We also recall how the existence of a conserved current does not correspond to a symmetry of the Lagrangian ${\cal L}$~\cite{Alexandre:2015oha}.
As we explain, a detailed study of the $\mathcal{PT}$-symmetry properties of the model is required in order
to understand its consistency. We then introduce in Sec.~\ref{sec:Path} a procedure for path-integral quantisation, 
which is based on the existence of a complete set of eigenstates with real energies in the $\mathcal{PT}$-symmetric phase of the model. We then introduce an extension of the concept of a saddle point and show that the integration of quantum fluctuations about this configuration is well defined. Finally, in Sec.~\ref{sec:Goldstone}, we discuss the extension of the Goldstone theorem
to the $\mathcal{PT}$-symmetric case, which follows the same steps as in an Hermitian theory, provided one considers 
$\mathcal{PT}$-conjugate instead of Hermitian-conjugate states. A summary and discussion of outstanding
issues are given in Sec.~\ref{sec:conx}.

\section{Complex scalar model}
\label{sec:model}

We consider a theory containing two complex scalar fields with the Lagrangian density
\be \label{lagrangian}
\mathcal{L}\ =\ \partial_\nu \phi_1^\star\, \partial^\nu \phi_1\: +\: \partial_\nu \phi_2^\star\, \partial^\nu \phi_2\: -\: m_1^2 |\phi_1|^2\: -\: m_2^2 |\phi_2|^2\: 
-\: \mu^2 \big( \phi_1^\star \phi_2\: -\: \phi_2^\star \phi_1 \big)\: -\:U_{\rm int}~,
\ee
in which the interaction potential $U_{\rm int}$ is $\mathcal{PT}$ symmetric.
The free part of this Lagrangian describes the simplest scalar model that contains a non-Hermitian but $\mathcal{PT}$-symmetric mass 
term \cite{Alexandre:2017foi}. The corresponding Hamiltonian is invariant under the combined action of the following $\mathcal{P}$ and $\mathcal{T}$ transformations:
\begin{subequations}
\begin{align}
\mathcal{P}:&\qquad \phi_1(t,\mathbf{x})\ \longrightarrow\ \phi_1'(t,-\,\mathbf{x})\ =\ +\:\phi_1(t,\mathbf{x})\;,\\
&\qquad \phi_2(t,\mathbf{x})\ \longrightarrow\ \phi_2'(t,-\,\mathbf{x})\ =\ -\:\phi_2(t,\mathbf{x})~,\\
\mathcal{T}:&\qquad \phi_1(t,\mathbf{x})\ \longrightarrow\ \phi_1'(-\,t,\mathbf{x})\ =\ \phi_1^{\star}(t,\mathbf{x})\;,\\
&\qquad \phi_2(t,\mathbf{x})\ \longrightarrow\ \phi_2'(-\,t,\mathbf{x})\ =\ \phi_2^{\star}(t,\mathbf{x})~.
\end{align}
\end{subequations}
Restricting our attention to the free part of the Lagrangian, it is convenient to introduce the doublet
\begin{equation}
\Phi(x)\ \equiv\ \begin{pmatrix} \phi_1(x) \\ \phi_2(x)\end{pmatrix}~.
\end{equation}
The $\mathcal{P}$ and $\mathcal{T}$ transformations can then be written in the condensed forms
\begin{subequations}
\begin{align}
\mathcal{P}:&\qquad \Phi(t,\mathbf{x})\ \longrightarrow\ \Phi'(t,-\,\mathbf{x})\ =\ P\Phi(t,\mathbf{x})~,\\
\mathcal{T}:&\qquad \Phi(t,\mathbf{x})\ \longrightarrow\ \Phi'(-\,t,\mathbf{x})\ =\ T\Phi^{\star}(t,\mathbf{x})~,
\end{align}
\end{subequations}
where $T\equiv{\rm diag}\,(1,1)$ and $P\equiv{\rm diag}\,(1,-1)$. We note that $\phi_1$ transforms as a scalar and $\phi_2$ transforms as a pseudoscalar. 

We can introduce the $\mathcal{PT}$ adjoint~\cite{Alexandre:2017foi} of $\Phi(x)$:
$\Phi^{\ddagger}(x)\equiv [\Phi^{\mathcal{PT}}(x)]^{\mathsf{T}}$, where the superscript $\mathsf{T}$ indicates the matrix transpose. 
Neglecting total derivatives (see below), the Lagrangian density can then be written as
\begin{equation}
\mathcal{L}\ =\ \Phi^{\ddagger}\begin{pmatrix} -\:\Box\:-\:m_1^2 & -\:\mu^2 \\ -\:\mu^2 & \Box\:+\:m_2^2\end{pmatrix}\Phi\:-\:U_{\rm int}~.
\end{equation}
The variation of the action due to variations in $\Phi$ and $\Phi^\ddagger$ is
\begin{align}
\label{eq:deltaS}
\delta S\ &=\ \int\!{\rm d}^4x\;\bigg[\bigg(\frac{\partial \mathcal{L}}{\partial \Phi}\:-\:\partial_{\nu}\,
\frac{\partial \mathcal{L}}{\partial(\partial_{\nu}\Phi)}\bigg)\delta\Phi\:+\:\delta \Phi^{\ddagger}\bigg(\frac{\partial \mathcal{L}}{\partial \Phi^{\ddagger}}\:
-\:\partial_{\nu}\,\frac{\partial \mathcal{L}}{\partial(\partial_{\nu}\Phi^{\ddagger})}\bigg)\nonumber\\&\qquad+\:\partial_{\nu}
\bigg(\frac{\partial \mathcal{L}}{\partial(\partial_{\nu}\Phi)}\,\delta \Phi\:+\:\delta \Phi^{\ddagger}\,\frac{\partial \mathcal{L}}{\partial(\partial_{\nu}\Phi^{\ddagger})}\bigg)
\bigg]~,
\end{align}
and we can quickly convince ourselves that the standard Euler-Lagrange equations
\begin{equation}
\frac{\partial \mathcal{L}}{\partial \Phi}\:-\:\partial_{\nu}\,\frac{\partial \mathcal{L}}{\partial(\partial_{\nu}\Phi)}\ =\ 0\qquad \text{and}\qquad 
\frac{\partial \mathcal{L}}{\partial \Phi^{\ddagger}}\:-\:\partial_{\nu}\,\frac{\partial \mathcal{L}}{\partial(\partial_{\nu}\Phi^{\ddagger})}\ =\ 0
\end{equation}
are inconsistent as a result of the non-Hermiticity. Thus, if we require $\delta S=0$, the support of non-trivial solutions $\Phi\neq 0$ would 
require the surface terms in the second line of Eq.~\eqref{eq:deltaS} to be non-vanishing. Alternatively, we must introduce an external source~\cite{Alexandre:2017foi}. 
Whichever course is taken, we can choose to fix the variational procedure with respect to either $\Phi$ or $\Phi^{\ddagger}$, i.e.~we can take
\begin{equation}
\frac{\delta S}{\delta \Phi}\ \equiv\ \frac{\partial \mathcal{L}}{\partial \Phi}\:-\:\partial_{\nu}\,\frac{\partial \mathcal{L}}{\partial(\partial_{\nu}\Phi)}\ 
=\ 0\qquad \text{or}\qquad \frac{\delta S}{\delta \Phi^{\ddagger}}\ \equiv\ \frac{\partial \mathcal{L}}{\partial \Phi^{\ddagger}}\:-\:\partial_{\nu}\,
\frac{\partial \mathcal{L}}{\partial(\partial_{\nu}\Phi^{\ddagger})}\ =\ 0~.
\end{equation}
Choosing the latter, the equations of motion are
\begin{subequations} \label{eq:eom}
\begin{gather}
\Box \phi_1\:+\:m_1^2\phi_1\:+\:\mu^2\phi_2\:+\frac{\partial U_{\rm int}}{\partial\phi_1^\star} =\ 0\;,\\
\Box \phi_2\:+\:m_2^2\phi_2\:-\:\mu^2\phi_1\:+\frac{\partial U_{\rm int}}{\partial\phi_2^\star}=\ 0\;.
\end{gather}
\end{subequations}
The squared mass eigenvalues
\begin{equation}
M^2_{\pm}\ =\ \frac{1}{2}\,\big(m_1^2\:+\:m_2^2\big)\:\pm\:\frac{1}{2}\sqrt{\big(m_1^2\:-\:m_2^2\big)^2\:-\:4\mu^4}
\end{equation}
are real so long as we remain in the region of unbroken $\mathcal{PT}$ symmetry, requiring
\be \label{Unbrokencondition}
\eta \ \equiv\ \frac{2 \mu^2}{|m_1^2\:-\: m_2^2|}\ \leq\ 1\;.
\ee

An additional consequence of the above subtlety in the variational procedure is the way in which conserved currents arise. 
Having chosen to define the variational procedure with respect to $\Phi^{\ddagger}$, a careful treatment of Noether's theorem 
(see Ref.~\cite{Alexandre:2017foi}) shows that there exists a conserved current for any transformation that satisfies
\begin{equation}
\label{eq:currentcond}
\delta \mathcal{L}\ =\ \bigg(\frac{\partial \mathcal{L}}{\partial \Phi}\:-\:\partial_{\nu}\,\frac{\partial \mathcal{L}}{\partial(\partial_{\nu}\Phi)}\bigg)\delta\Phi\ 
=\ 2\mu^2\big(\phi_2^{\star}\delta \phi_1\:-\:\phi_1^{\star}\delta \phi_2\big)\:-\:2i \left[ \frac{\partial}{\partial\Phi}\, {\rm Im}\, U_{\rm int} \right] \delta\Phi~,
\end{equation}
Notice that $\delta \mathcal{L}=0$ in the Hermitian limit, and we recover the usual statement of Noether's theorem~\cite{Noether}: 
for every continuous symmetry of the Lagrangian, there exists a corresponding conserved current. This is not the case for our non-Hermitian theory. The two $U(1)$ currents
\begin{equation}
j_1^{\nu}\ =\ i\big(\phi_1^{\star}\partial^{\nu}\phi_1\:-\:\phi_1\partial^{\nu}\phi_1^{\star}\big)\qquad \text{and}\qquad 
j_2^{\nu}\ =\ i\big(\phi_2^{\star}\partial^{\nu}\phi_2\:-\:\phi_2\partial^{\nu}\phi_2^{\star}\big)
\end{equation}
are not conserved in the free theory for $\mu\neq 0$; specifically,
\begin{equation}
\partial_{\nu}j_1^{\nu}\ =\ \partial_{\nu}j^{\nu}_2\ =\ i\mu^2\big(\phi_2^{\star}\phi_1\:-\:\phi_1^{\star}\phi_2\big)~.
\end{equation}
Their difference $j^{\nu}\equiv j_1^{\nu}-j_2^{\nu}$, however, is conserved, and this current corresponds to the $U(1)$ transformations
\begin{subequations} \label{eq:contTransf}
\begin{align}
\phi_1(x)\ &\longrightarrow\ \phi_1'(x)\ =\ e^{+i\epsilon}\phi_1(x)~,\\
\phi_2(x)\ &\longrightarrow\ \phi_2'(x)\ =\ e^{-i\epsilon}\phi_2(x)~,
\end{align}
\end{subequations}
which satisfy Eq.~\eqref{eq:currentcond} but do \emph{not} leave the Lagrangian invariant. In fact, these transformations yield a one-parameter family of \emph{equivalent} non-Hermitian theories, whose free Lagrangians have the form
\begin{equation}
\label{eq:Lalpha}
\mathcal{L}_{\epsilon}\ =\ \partial_\nu \phi_1^\star\, \partial^\nu \phi_1\: +\: \partial_\nu \phi_2^\star\, \partial^\nu \phi_2\: -\: m_1^2 |\phi_1|^2\: -\: m_2^2 |\phi_2|^2\: 
-\: \mu^2 e^{-2i\epsilon} \phi_1^\star \phi_2\: +\: \mu^2\,e^{+2i\epsilon}\phi_2^\star \phi_1
\end{equation}
and whose mass spectra are identical. That is to say, whilst the Lagrangian is not invariant under the transformations associated with the conserved current, physical quantities, such as the masses, are. Finally, we note that comments on the non-trivial relation between symmetries and conservation laws in non-Hermitian quantum mechanics can be found 
in Ref.~\cite{Muga}.

\section{Path-integral formulation}
\label{sec:Path}

We now turn out attention to the formulation of the path-integral representation of the non-Hermitian field theory.

\subsection{New conjugate field variables}

The Lagrangian in Eq.~\eqref{lagrangian} would naively appear to have a finite imaginary part for $\mu\neq 0$, and one might be concerned that this
could modify the convergence of the path integral. However, the spectrum of this theory is real and positive definite in the region 
of unbroken $\mathcal{PT}$ symmetry, enabling us to formulate consistently the path integral and its quantisation.

We can rotate to the mass eigenbasis via the transformation
\begin{equation}
\Xi\ \equiv\ R \Phi\ =\ \begin{pmatrix}
	\xi_1 \\ \xi_2
\end{pmatrix}~,\qquad
\bar{\Xi}\ \equiv\ \Phi^{\dag}R^{-1}\ =\ \begin{pmatrix}
	\bar{\xi}_1 \\ \bar{\xi}_2
\end{pmatrix}~,
\end{equation}
where
\be \label{Rmatrix}
R \ =\ \mathcal{N}\begin{pmatrix}
	\eta &  1 - \sqrt{1 - \eta^2}  \\1 - \sqrt{1 - \eta^2}   & \eta
\end{pmatrix}\;,
\ee
with
\begin{equation}
\mathcal{N}^{-1}\ \equiv \ \sqrt{2 \eta^2\: -\: 2\: +\: 2\sqrt{1 \:-\: \eta^2}}~.
\end{equation}
The matrix $R$ satisfies the following properties
\be
R^\dagger\ =\ R~,\qquad R^{-1}\ =\ PRP^{-1}\ =\ PRP~,
\ee
such that 
\be
\bar\Xi \ =\ \Xi^\ddagger C'~,\qquad \mbox{with}\qquad \Xi^\ddagger\ =\ \Xi^\dagger P~,\qquad C'\ =\ RPR^{-1}~.
\ee
The variables $\Xi$ and $\bar{\Xi}$ are $\mathcal{C'PT}$-conjugate fields in the sense of Ref.~\cite{Bender:2005tb}. 
We note that the $\mathcal{C}'$ transformation here, which we identify with a prime, is not the canonical $\mathcal{C}$ transformation in Fock space, 
which would involve complex conjugation. 
Instead, it is the transformation by which one constructs the positive-definite inner product in $\mathcal{PT}$-symmetric quantum mechanics~\cite{Bender:2002vv} (see also Ref.~\cite{Bender:2005tb}).

The free Lagrangian becomes
\begin{equation}
\mathcal{L}_0\ =\ \bar{\Xi}\,\Delta^{-1}\,\Xi~,\qquad\mbox{where}\qquad\Delta^{-1}\ =\ 
\begin{pmatrix}
	-\:\Box\: -\: M_+^2 & 0 \\ 0 & -\:\Box\: -\: M_-^2
\end{pmatrix}~,
\end{equation}
and it appears to be that of an Hermitian theory. However, introducing interactions leads to the non-trivial feature mentioned above: 
varying the full action with respect to 
$(\xi_1,\xi_2)$ or $(\bar\xi_1,\bar\xi_2)$ does not yield the same equations of motion.
This can be seen, for example, with the interaction $|\phi_1\phi_1^\star|^2$, which can be expressed using either $\Phi=R^{-1}\Xi$:
\begin{equation}
|\phi_1\phi_1^\star|^2\ =\ |\phi_1|^4\ =\ \mathcal{N}^4\big|\eta\,\xi_1\: +\: \big(\sqrt{1\: -\: \eta^2}\: -\: 1\big) \xi_2\big|^4~,
\end{equation}
or $\Phi^\dagger=\bar\Xi R$:
\be
|\phi_1\phi_1^\star|^2\ =\ |\phi_1^\star|^4\ =\ \mathcal{N}^4\big|\eta\,\bar\xi_1\: -\: \big(\sqrt{1\: -\: \eta^2}\: -\: 1\big) \bar\xi_2\big|^4~.
\ee

\subsection{Partition function}

The partition function is obtained from the vacuum persistence amplitude in the presence of external sources
\be 
J \ =\ \begin{pmatrix} j_1 \\ j_2 \end{pmatrix}\qquad\mbox{and}\qquad\bar J\ =\ J^\ddagger C'~.
\ee
For the non-Hermitian theory, this vacuum persistence amplitude is
\begin{equation}
Z[J,\bar{J}]\ =\ \braket{\bar{0}(+\infty)|0(-\infty)}_{J,\bar{J}}~,
\end{equation}
where the state $\bra{\bar{0}}$ is the $\mathcal{C'PT}$ conjugate of the vacuum state. The path integral is developed in the usual way, 
except that one must insert complete sets of eigenstates of the Heisenberg-picture field operator $\Xi$ and its $\mathcal{C'PT}$ conjugate $\bar{\Xi}$ 
(rather than its Hermitian conjugate) at all intermediate times. In this way, one arrives at the following result for the Euclidean path integral:
\be \label{partfct}
Z[J,\bar{J}] \ =\  \int \mathcal{D} [\Xi,\bar{\Xi}]\exp\left( {-\:S_E[\Xi,\bar{\Xi}]\: +\int_x\; \big( \bar{J}\,\Xi \:+\: \bar{\Xi}\,J \big)}\right)~,
\ee
where $S_E$ is the Euclidean action and we use the shorthand notation \smash{$\int_x\equiv\int{\rm d}^4x$}. Of course, having established the correct form for the partition function, one could rewrite it in terms of the original $\mathcal{PT}$-conjugate variables $\Phi$ 
and $\Phi^{\ddagger}$ by making the change of variables and accounting for the functional Jacobian, which is non-trivial but field independent.

The partition function (\ref{partfct}) can be expanded around the free part
\bea
Z[J,\bar J]\ &=&\ \int \mathcal{D} [\Xi,\bar{\Xi}]\;\exp\Bigg[-\int_x \bar\Xi\Delta^{-1}\Xi+\int_x (\bar J\Xi+\bar\Xi J)-\int_x U_{\rm int}\Bigg]\nonumber\\
&=&\ \exp\Bigg[\int_x\bar J\Delta J\Bigg]\int \mathcal{D}[\Pi,\bar\Pi]\;\exp\Bigg[-\int_x\bar\Pi\Delta^{-1}\Pi-\int_x U_{\rm int}\Bigg]\nonumber\\
&=&\ \exp\Bigg[\int_x\bar J\Delta J\Bigg]\sum_{n\,=\,0}^\infty\frac{1}{n!}
\int \mathcal{D}[\Pi,\bar\Pi]\;\exp\Bigg[-\int_x\bar\Pi\Delta^{-1}\Pi\Bigg]\Bigg[-\int_x U_{\rm int}\Bigg]^n~,
\eea
where $\Delta^{-1}=\mathrm{diag}\,(-\,\partial^2+M_+^2,-\,\partial^2+M_-^2)$ in Euclidean signature and $\Pi\equiv\Xi-\Delta J=(\pi_1\;, \pi_2)^{\mathsf{T}}$. One can see that the perturbative structure is the usual one, comprising well-defined Gaussian integrals at each order.

\subsection{One-loop 1PI effective action}

There is an unambiguous definition of the classical saddle point $(\Xi_0,\bar\Xi_0)$ for the path integral (\ref{partfct}), which satisfies
\be\label{equamotXi0}
\bigg[-\:\frac{\delta S_E}{\delta\Xi}\:+\:\bar J\bigg]_{\!0}\ =\ 0\ =\ \bigg[-\:\frac{\delta S_E}{\delta\bar\Xi}\:+\:J\bigg]_{\!0}~,
\ee
where the index 0 indicates evaluation at the configuration $(\Xi_0,\bar\Xi_0)$.
Expanding the partition function up to quadratic order around the saddle point, we obtain for the one-loop partition function 
\bea\label{Z1}
Z^{(1)}[J, \bar J] \ &=&\ \exp \Bigg[ -S_E[\Xi_0, \bar\Xi_0]\: +\: \int_x \Big( \bar J\Xi_0+\bar\Xi_0 J \Big)\Bigg] \nonumber\\
&&\qquad \times\:\int \mathcal{D}[\Xi,\bar\Xi]\;\exp \Bigg[ -\:\frac{1}{2}\int_{xy} \bigg(2\left(\bar\Xi - \bar\Xi_0\right)_x
\frac{\delta^2 S_E}{\delta \bar\Xi_x \delta \Xi_y}\Bigg|_0 \left(\Xi - \Xi_0\right)_y \nonumber\\&&\qquad +\:\left(\bar\Xi - \bar\Xi_0\right)_x
\frac{\delta^2 S_E}{\delta \bar\Xi_x \delta \bar\Xi_y}\Bigg|_0 \left(\bar\Xi - \bar\Xi_0\right)_y\:+\:\left(\Xi - \Xi_0\right)_x
\frac{\delta^2 S_E}{\delta \Xi_x \delta \Xi_y}\Bigg|_0 \left(\Xi - \Xi_0\right)_y\bigg)\Bigg] \nonumber \\ 
&=&\ \exp \Bigg[ -\:S_E[\Xi_0, \bar\Xi_0]\: +\: \int_x \Big( \bar J \Xi_0+\bar\Xi_0 J \Big) \:
-\: \frac{1}{2}\, \text{STr}\, \ln S^{(2)}_E\Big|_0\Bigg]~,
\eea
where $S^{(2)}_E$ is the functional Hessian matrix (in field space) of the Euclidean action and $\text{STr}$ indicates the trace over both coordinate and field spaces. In order to define the one-particle irreducible (1PI) effective action $\Gamma^{(1)}$, one introduces the background field $\Xi_c$:
\be\label{xic}
\Xi_c\ =\ \frac{1}{Z^{(1)}}\,\frac{\delta Z^{(1)}}{\delta \bar J}~,
\ee
which, from Eq.~\eqref{Z1}, is 
\bea
\Xi_c\ &=&\ \Xi_0\:+\:\int_x\left(-\:\frac{\delta S_E}{\delta\Xi_0}\:+\:\bar J\right)\frac{\delta\Xi_0}{\delta \bar J}\:
-\:\frac{1}{2}\,\frac{\delta}{\delta\bar J}\,\,\text{STr}\,\ln S^{(2)}_E\Big|_0 \nonumber \\
&=&\ \Xi_0\:+\:\text{quantum corrections}~.
\eea
$\Gamma^{(1)}$ is then defined after inverting the relation (\ref{xic}) to express $\bar J$ as a functional of $\Xi_c$:
\bea
\Gamma^{(1)}[\Xi_c,\bar\Xi_c]\ &=&\ -\:\ln Z^{(1)}\:+\:\int_x\Big( \bar J \Xi_c+\bar\Xi_c J \Big)\nonumber\\
&=&\ S_E[\Xi_c,\bar\Xi_c]\:+\:\frac{1}{2}\,\text{STr}\,\ln S^{(2)}_E\Big|_c~,
\eea
were the index $c$ indicates evaluation in the background field configuration.
The one-loop 1PI effective potential is obtained for a constant configuration $\Xi_c$ and is then given by 
\be
U^{(1)}(\Xi_c,\bar\Xi_c) \ =\ U(\Xi_c,\bar\Xi_c)\: +\: \frac{1}{2 V^{(4)}}\, \text{STr}\,\ln S^{(2)}_E\Big|_c~,
\ee
where $V^{(4)}$ is the spacetime volume. After a rotation to the original basis, which does not affect the trace, we finally obtain
\be\label{1PIpot}
U^{(1)}(\Phi_c,\Phi_c^\dagger)\ =\ 
U(\Phi_c,\Phi_c^\dagger)\: +\: \frac{1}{2V^{(4)}}\,\text{STr}\,\ln S^{(2)}_E\Big|_c~.
\ee

\subsection{Running couplings}

We consider here a bare interaction potential of the form 
\bea\label{BarePot}
U_{\rm int}^{(0)} \ &=&\ 
\frac{g_1}{4} |\phi_1|^4 \:+\: \frac{g_2}{4} |\phi_2|^4 
\:+\: \lambda |\phi_1 \phi_2|^2\:+\: \frac{\alpha}{4}\, \Big(\left( \phi_1^\star \phi_2 \right)^2\: +\: \left( \phi_2^\star \phi_1 \right)^2 \Big)\\&&\qquad 
+\:\frac{1}{2}\, \Big(  \beta_1 |\phi_1|^2 + \beta_2 |\phi_2|^2 \Big)\Big( \phi_1^\star \phi_2 - \phi_2^\star \phi_1 \Big)~.\nonumber
\eea
Substituting this potential into Eq.~\eqref{1PIpot} leads to the following one-loop running of the coupling constants
(details can be found in the Appendix):
\begin{subequations}
\label{oneloopcorrections}
\bea
\left( m_i^{2} \right)^{(1)} \: &=&\:  m_i^{2}\:  +\: \frac{g_i + \lambda }{16 \pi^2}\Lambda^2\: +\: \mathcal{O}\left(\ln\left(\frac{\Lambda}{m}\right)\right)\;, \\ 
\left( \mu^2 \right)^{(1)} \: &=&\: \mu^2\: +\: \frac{\beta_1 + \beta_2 }{16 \pi^2}\Lambda^2\: -\: \frac{1}{8\pi^2} \Big( \mu^2 (\lambda - \alpha) 
+ \beta_1 m_1^2 + \beta_2 m_2^2 \Big) \ln\left( \frac{\Lambda}{m} \right)\;, \\ 
g_i^{(1)} \: &=&\:  g_i\: -\: \frac{1}{16\pi^2}\Big(  5 g_i^2 + \alpha^2 + 4 \lambda^2 - 10 \beta_i^2  \Big) \ln \bigg( \frac{\Lambda}{m} \bigg)\;, \\
\lambda^{(1)} \: &=&\: \lambda\: -\: \frac{1}{16\pi^2}\Big( 4 \lambda^2 + 2 \alpha^2 + 2 \lambda\left( g_1 + g_2 \right) - 3 \left( \beta_1^2 + \beta_2^2 \right) - 4 \beta_1 \beta_2 \Big) 
\ln \bigg( \frac{\Lambda}{m} \bigg)\;, \\
\alpha^{(1)} \: &=&\: \alpha\: -\: \frac{1}{16\pi^2}\Big( 4 \left( \beta_1^2 + \beta_2^2 \right) + \alpha \left( g_1 + g_2 \right) + 2 \beta_1 \beta_2 + 8 \lambda \alpha \Big) 
\ln \bigg( \frac{\Lambda}{m} \bigg)\;,\\ 
\beta_i^{(1)} \: &=&\:  \beta_i\: -\: \frac{1}{16\pi^2}\Big( 5 g_i \beta_i + 4 \beta_j \lambda - \alpha \beta_j + 6 \lambda \beta_i - 4 \alpha \beta_i \Big) 
\ln \bigg( \frac{\Lambda}{m} \bigg)\;,
\eea
\end{subequations}
where $m$ is a typical mass scale of the system, $i\ne j$, and finite terms are omitted.

\subsection{Hermitian fixed point}

We assume here that the non-Hermitian interactions are switched off ($\beta_i=0$) and the only source of non-Hermiticity is the mass parameter $\mu^2$.  
Quantum corrections modify this mass parameter, and we need to check that the condition (\ref{Unbrokencondition}), which 
delineates the phase of unbroken $\mathcal{PT}$ symmetry,
remains valid at one loop. For a fixed set of dressed parameters, the one-loop running of the parameter $\eta$ is 
\be
\eta(\Lambda)\ =\
\Bigg| \frac{2(\mu^2)^{(1)}  - (\alpha^{(1)} - 
\lambda^{(1)})\mu^2 /(4 \pi^2)\ln\left( \Lambda / m \right))}{(m_1^2)^{(1)} - (m_2^2)^{(1)} - \Big( g_1^{(1)} - g_2^{(1)} \Big) \Lambda^2 /(16 \pi^2)} \Bigg|~.
\ee
We recall that, for the $\mathcal{PT}$ symmetry to be unbroken, the following requirement needs to be satisfied for all values of $\Lambda$:
\be 
\eta(\Lambda) \ <\ 1~.
\ee
If $g_1^{(1)}\ne g_2^{(1)}$, we can see that $\eta(\Lambda)\to0$ when $\Lambda\to\infty$, such that the theory converges to a 
Hermitian limit, which thus appears as an UV fixed point.

\section{Goldstone modes}
\label{sec:Goldstone}

Having established a consistent formulation of the non-Hermitian path integral and its quantisation,
we show, in this section, that the usual proof for the presence of Goldstone modes is still valid in the $\mathcal{PT}$-symmetric case, and we explicitly derive these modes at one-loop order. 
As explained below, the existence of a Goldstone mode relies on a conserved current and not on the invariance of the Lagrangian. We note, however, that 
both are related: in the model (\ref{lagrangian}), current conservation arises from the field transformation $\Phi\to\exp(i\epsilon P)\Phi$, whereas the 
Lagrangian is invariant under the transformation $\Phi\to\exp(i\epsilon)\Phi$. The Goldstone mode is a consequence of the former transformation, but
the choice of a specific vacuum spontaneously breaks the latter symmetry.

\subsection{Proof of the Goldstone theorem}

Before considering our specific example, we first revisit the derivation of the Goldstone theorem~\cite{GSW1,GSW2,GSW3} in the context of a non-Hermitian theory. 
We assume that there exists an infinitesimal transformation, which takes the generic form
\begin{equation}
\Phi\ \to \ \Phi\:+\:i\epsilon T\Phi~,
\end{equation}
where $T$ is the generator of the transformation. We also assume that this transformation corresponds to a conserved current $j^{\nu}$ with conserved charge 
$Q=\int{\rm d}^3\mathbf{x}\;j^0(x)$. Most importantly, for the non-Hermitian theory, this transformation does not leave the Lagrangian invariant.

We are interested in the vacuum expectation of the commutator $[ Q, \Phi(x)]$:
\be 
\braket{\bar{0}|[ Q, \Phi(x)]|0}\ =\ i T\braket{\Phi}~,
\ee
where $\braket{\Phi}\equiv\braket{\bar{0}|\Phi(x)|0}$. We note that the inner product is defined with respect to $\mathcal{C}'\mathcal{PT}$, as is necessary 
for a non-Hermitian theory. With this exception, the proof of the Goldstone theorem proceeds in the same manner as for Hermitian theories (and we closely follow Ref.~\cite{EWeinbergNotes}). By inserting complete 
sets of intermediate states, we can write
\begin{align}
\langle \bar{0} | [ j^\nu(y), \Phi(x) ] | 0 \rangle \ &=\ \sum_N \Big[\langle \bar{0} | j^\nu(y) | N \rangle \langle \bar{N} | \Phi(x) | 0 \rangle\: -\: \langle \bar{0} | \Phi(x) 
| N \rangle \langle \bar{N} | j^\nu(y) | 0 \rangle\Big] \nonumber  \\ 
&= \ \int\!\frac{{\rm d}^4 p}{(2\pi)^4}\;e^{-ip\cdot(y-x)}\sum_N\Big[(2\pi)^4\delta^4(p_N-p)\langle \bar{0} | j^\nu(0) | N \rangle \langle \bar{N} | \Phi(0) | 0 \rangle \nonumber\\&\qquad-\:(2\pi)^4\delta^4(p_N+p)\langle \bar{0} | \Phi(0) | N \rangle \langle \bar{N} | j^\nu(0) | 0 \rangle \Big]~,
\end{align}
and, by virtue of Lorentz invariance, we have that
\begin{subequations}
\label{eq:Goldint}
\bea 
\sum_N (2\pi)^4\delta^4(p_N-p) \langle \bar{0} | j^\nu(0) | N \rangle \langle \bar{N} | \Phi(0) | 0 \rangle \ &=&\ 2\pi i \theta(+\,p_0)p^\nu \rho(p^2)~, \\
\sum_N (2\pi)^4\delta^4(p_N+p) \langle \bar{0} | \Phi(0) | N \rangle \langle \bar{N} | j^\nu(0) | 0 \rangle \ &=&\ 2\pi i\theta(-\,p_0) p^\nu \bar{\rho}(p^2)~.
\eea
\end{subequations}
Moreover, causality requires that the commutator vanish for space-like separations, and it follows that $\rho(p^2)=\bar{\rho}(p^2)$. We then arrive at the (K\"{a}ll\'{e}n-Lehmann) spectral representation
\bea 
\langle \bar{0} | [j^\nu(y), \Phi(x)] |0 \rangle\ &=&\ i\!\int\!\frac{{\rm d}^4 p}{(2\pi)^4}\; e^{-ip\cdot(y-x)}\,2\pi\,{\rm sgn}(p_0)\,p^{\nu}\rho(p^2) \nonumber  \\ 
&=& \ -\:\frac{\partial}{\partial y_\nu} \int\!{\rm d} \sigma^2\; \rho(\sigma^2) \Delta(y,x;\sigma^2)~,
\eea
where
\begin{equation}
\Delta(y,x;\sigma^2)\ =\ \int\!\frac{{\rm d}^4p}{(2\pi)^4}\;e^{-ip\cdot (y-x)}\;2\pi\,{\rm sgn}(p_0)\,\delta(p^2-\sigma^2)~,
\end{equation}
is the Pauli-Jordan function with the mass of the field replaced by $\sigma$.

Since the current is conserved, it follows that
\bea
-\:\Box_y \int\! {\rm d} \sigma^2\;\rho(\sigma^2) \Delta(y,x;\sigma^2)\ =\ \int\! {\rm d} \sigma^2\;\sigma^2\rho(\sigma^2) \Delta(y,x;\sigma^2)\ =\ 0~,
\eea
in which case $\rho(\sigma^2)$ must be zero for $\sigma^2\neq 0$, i.e.~$\rho(\sigma^2)=\rho_0\delta(\sigma^2)$. Thus, for $x_0=y_0$, we have
\bea 
\langle \bar{0} | [j^0(y), \Phi(x)] |0 \rangle\ &=&\ i\rho_0\,\delta^3(\mathbf{y}-\mathbf{x})~,
\eea
and it follows that
\begin{equation}
\braket{\bar{0}|[ Q, \Phi(x)]|0}\ =\ i T \braket{\Phi}\ =\ i\rho_0~.
\end{equation}
If there exists a non-trivial vacuum $\braket{\Phi}$, which is not invariant under the transformation generated by $T$, then $\rho_0\neq 0$. 
We remark that, for a non-Hermitian theory, $\braket{\Phi}'= T\braket{\Phi}$ is a vacuum state of the \emph{transformed} Lagrangian, e.g., for the transformations in Eq.~\eqref{eq:contTransf}, $\braket{\Phi}'$ is the vacuum state with respect to the Lagrangian in Eq.~\eqref{eq:Lalpha}.
The latter fact does not, however, affect the derivation of the Goldstone theorem. Returning to the expressions in Eq.~\eqref{eq:Goldint}, we have
\be 
\sum_N (2\pi)^4\delta^4(p_N-p) \langle \bar{0} | j^\nu(0) | N \rangle \langle \bar{N} | \Phi(0) | 0 \rangle \ =\ 2\pi i\theta(+\,p_0) p^\nu \rho_0\delta(p^2)~.
\ee 
The right-hand side is non-vanishing when $p^2=0$, provided $p^{\nu}\neq 0^{\nu}$. It follows that there must exist a state $\ket{N}$ with $p_N=p$, such that $p_N^2=0$, 
i.e.~there must exist a massless state. 

We emphasise that this proof of the existence of a massless Goldstone mode relies on the existence of a conserved current and not on invariance of the Lagrangian. 
Hence, the Goldstone theorem persists for the non-Hermitian theory, and we give further details for our specific model in what follows.

\subsection{Spontaneous symmetry breaking}

In order to study spontaneous symmetry breaking, we consider the Lagrangian (\ref{lagrangian}) with $U_{\rm int}=g |\phi_1|^4/4$ and change the sign of the $m_1^2$ 
mass term, i.e.
\be 
\mathcal{L}\ =\ \partial_\nu \phi_1^\star \partial^\nu \phi_1 \:+\: \partial_\nu \phi_2^\star \partial^\nu \phi_2\: +\: m_1^2 |\phi_1|^2\: -\: m_2^2 |\phi_2|^2 
\:-\: \mu^2 \big( \phi_1^\star \phi_2 - \phi_2^\star \phi_1 \big) \:-\: \frac{g}{4} |\phi_1|^4~,
\ee
which allows for a non-trivial vacuum structure. The vacuum expectation values are the solutions of the equations
\begin{subequations}
\bea 
\frac{\delta U}{\delta \phi_1^\star} \ &=&\ \frac{g}{2}|\phi_1|^2 \phi_1\: -\: m_1^2 \phi_1 \: +\: \mu^2 \phi_2 \ =\  0 ~,\\
\frac{\delta U}{\delta \phi_2^\star} \ &=&\ m_2^2 \phi_2\: -\: \mu^2 \phi_1 \ =\  0~,
\eea
\end{subequations}
conditions similar to the equations of motion. These equations are invariant under a phase transformation acting identically on both the fields. 
The non-trivial solutions to these equations are given by 
\be \label{vacuum}
 \begin{pmatrix}
 	v_1 \\ v_2
 \end{pmatrix} \ =\  \sqrt{2\,\frac{ m_1^2 m_2^2 - \mu^4}{g m_2^2}} \begin{pmatrix}
 1 \\ \frac{\mu^2}{m_2^2}
\end{pmatrix} e^{i \epsilon}~.
\ee  
For a fixed phase $\epsilon$, we can express the fields as fluctuations around these vacua
\be
\phi_1 \ =\  v_1\: +\: \hat{\phi}_1\qquad\mbox{and}\qquad 
\phi_2\ =\ v_2\: +\: \hat{\phi}_2~.
\ee
Expressing the equations of motion (\ref{eq:eom}) in terms of the field fluctuations $\hat{\phi}_{1,2}$ gives 
\be \label{eomInFluctuations}
\Box \begin{pmatrix}
\hat{\phi_1} \\ \hat{\phi}_1^\star \\ \hat{\phi}_2 \\ \hat{\phi}_2^\star
\end{pmatrix} \ =\  \begin{pmatrix}
\frac{\left( m_1^2 m_2^2 - 2 \mu^4 \right)}{m_2^2} & \frac{\left( m_1^2 m_2^2 -  \mu^4 \right)}{m_2^2} & \mu^2 & 0 \\ 
\frac{\left( m_1^2 m_2^2 -  \mu^4 \right)}{m_2^2} & \frac{\left( m_1^2 m_2^2 - 2 \mu^4 \right)}{m_2^2} & 0 & \mu^2 \\ 
-\mu^2 & 0 & m_2^2 & 0 \\ 
0 & -\mu^2 & 0 & m_2^2
\end{pmatrix} \begin{pmatrix}
\hat{\phi}_1 \\ \hat{\phi}_1^\star \\ \hat{\phi}_2 \\ \hat{\phi}_2^\star
\end{pmatrix} \:+\: \cdots~,
\ee
where the dots represent terms of higher order in $\hat{\phi}_1$ and $\hat{\phi}_1^\star$. 

It is easy to check that the determinant of this mass matrix is zero, and we therefore have the anticipated Goldstone mode. We remark that, 
whilst the explicit forms of the eigenmodes depend on the choice of the equations of motion, the eigenspectrum is unique.

The mass matrix has a single zero eigenvalue, and the corresponding (Goldstone) mode is
\be 
G_1 \ =\  \sqrt{\frac{2 m_2^4}{m_2^4 + \mu^4}} \bigg( \text{Im}\, \hat{\phi}_1\: -\: \frac{\mu^2}{m_2^2}\, \text{Im}\,  \hat{\phi}_2 \bigg)~.
\ee
For completeness, we list the other eigenvalues and their corresponding eigenmodes:
\begin{subequations}
\bea 
\lambda_2 \ &=&\ m_2^2\:-\:\frac{\mu^4}{m_2^2}~,\\ 
\lambda_3 \ &=&\ \frac{1}{2m_2^2}\,\bigg( 2m_1^2 m_2^2 \:-\: 3\mu^4 \:+\: m_2^4 \:+\: \sqrt{\left(2m_1^2m_2^2-3\mu^4 - m_2^4\right)^2-4\mu^4 m_2^4}\,\bigg)~,\\ 
\lambda_4 \ &=&\ \frac{1}{2m_2^2}\,\bigg( 2m_1^2 m_2^2\: -\: 3\mu^4 \:+\: m_2^4\: -\: \sqrt{\left(2m_1^2m_2^2-3\mu^4 - m_2^4\right)^2-4\mu^4 m_2^4} \bigg)~,
\eea
\end{subequations}
with
\begin{subequations}
\bea
G_2 \ &=&\ \sqrt{\frac{2 m_2^4}{m_2^4 + \mu^4 }}\,\bigg( \text{Im} \,\hat{\phi}_2\:-\: \frac{\mu^2}{m_2^2} \,\text{Im}\, \hat{\phi}_1 \bigg)~,\\
G_3\ &=&\ \sqrt{2}\,\bigg[1 + \bigg( \frac{\mu^2}{\lambda_3-m_2^2} \bigg)^2\, \bigg]^{-1/2} \bigg[ \text{Re}\, \hat{\phi_1}\: 
+\:\bigg( \frac{\mu^2}{\lambda_3-m_2^2}\bigg)\text{Re}\,\hat{\phi}_2 \bigg]~, \\ 
G_4\ &=&\ \sqrt{2}\,\bigg[ 1+\bigg( \frac{\lambda_4 - m_2^2}{\mu^2} \bigg)^2\,\bigg]^{-1/2}\bigg[ \text{Re}\,\hat{\phi}_2\: 
+\: \bigg( \frac{\lambda_4 - m_2^2}{\mu^2} \bigg)\text{Re}\,\hat{\phi}_1\bigg]~.
\eea
\end{subequations}
The form of the Goldstone mode could also have been anticipated from the conserved current itself. The conservation equation yields
\begin{equation}
\partial_{\nu}j^{\nu}\ =\ i\partial_{\nu}\big[\big(\phi_1^{\star}\partial^{\nu}\phi_1\:-\:\phi_1\partial^{\nu}\phi_1^{\star}\big)\:-
\:\big(\phi_2^{\star}\partial^{\nu}\phi_2\:-\:\phi_2\partial^{\nu}\phi_2^{\star}\big)\big]\ =\ 0~.
\end{equation}
Expanding this to first order in the fluctuations [setting the constant phase in the vacuum expectation values (vevs) $v_1$ and $v_2$ to zero] gives
\begin{equation}
\partial_{\nu}j^{\nu}\ \simeq\ -\:2\big(v_1\,\Box\,{\rm Im}\,\hat{\phi}_1\:-\:v_2\,\Box\,{\rm Im}\,\hat{\phi}_2\big)~,
\end{equation}
and we see that the Goldstone mode is
\begin{equation}
G_1\ \propto\ {\rm Im}\,\hat{\phi}_1\:-\:\frac{\mu^2}{m_2^2}\,{\rm Im}\,\hat{\phi}_2~.
\end{equation}
Finally, we note that for our choice of equations of motion, the Goldstone mode is in fact the left eigenvector of the mass matrix (as dictated by the conserved current).
Choosing the alternative definition of the variational procedure, the Goldstone mode would instead correspond to the right eigenvector of the mass matrix in Eq.~\eqref{eomInFluctuations}, 
which is distinct and related to the previous one by $\mathcal{PT}$ conjugation. Note that this is consistent with $\mathcal{PT}$ transformation superseding Hermitian conjugation for non-Hermitian theories and that the alternative definitions are equivalent.

\subsection{The Goldstone mode to one-loop order}

The full tree-level potential is given in terms of the fields $\hat{\phi}_1$ and $\hat{\phi}_2$ as
\bea 
U^{(0)} \ &=&\ M_1^2 |\hat{\phi}_1|^2\: +\: m_2^2 \hat{\phi}_2 \left( \hat{\phi}_2^\star + M_c \right) 
\:+\: \mu^2 \Big( \hat{\phi}_2 \hat{\phi}_1^\star - \hat{\phi}_1 \left( \hat{\phi}_2^\star + M_c \right) \Big) \nonumber \\
&&\qquad +\: \frac{M_a^2}{2} \left( \hat{\phi}_1^2 \:+\: \left(\hat{\phi}_1^\star \right)^2 \right) 
\:+\: \frac{M_b}{2}|\hat{\phi}_1|^2 \left( \hat{\phi}_1^\star + \hat{\phi}_1 \right) \:+\: \frac{g}{4}\,|\hat{\phi}_1|^4~,
\eea
where we use the notation 
\begin{subequations}
\bea
M_1^2\ &=&\ \frac{m_1^2m_2^2 - 2\mu^4}{m_2^2} ~,\\ 
M_a^2\ &=&\ \frac{m_1^2 m_2^2 - \mu^4}{ m_2^2}\ =\ M_1^2 \:+\: \frac{\mu^4}{m_2^2}~,\\  
M_b\ &=&\ g \sqrt{2\,\frac{m_1^2m_2^2-\mu^4}{g m_2^2}}~,\\
M_c \ &=&\ \frac{2 \mu^2}{m_2^2}\, \sqrt{2 \frac{m_1^2m_2^2-\mu^4}{gm_2^2}} ~.
\eea
\end{subequations}
The linear terms in the potential are a consequence of the non-Hermitian nature of the system. 
At one-loop level, these couplings are obtained by 
substituting this potential into Eq.~\eqref{1PIpot} and are given by 
\begin{subequations}
\bea 
g^{(1)} \ &=&\ g \:-\: \frac{5 g^2}{16 \pi^2} \ln \bigg(\frac{\Lambda}{m} \bigg) ~,\\
m_2^{2(1)} \ &=&\ m_2^2 ~,\\ 
\mu^{2(1)} \ &=&\ \mu^2 ~,\\
M_1^{2(1)} \ &=&\ M_1^2 \:+\: \frac{g \Lambda^2}{16 \pi^2} \:+\: \mathcal{O}\left( \ln \left( \Lambda/m \right) \right)~,\\ 
M_a^{2(1)} \ &=&\ M_a^2 \:-\: \frac{1}{16\pi^2}\left( 2 M_b^2 + g M_a^2 \right) \ln \bigg(\frac{\Lambda}{m}\bigg) ~,\\ 
M_b^{(1)} \ &=&\ M_b\: -\: \frac{5gM_b }{16 \pi^2}\ln \bigg( \frac{\Lambda}{m} \bigg) ~,
\eea
\end{subequations}
where finite terms are again omitted.
A linear term is also generated, which is given by 
\be 
M_b \frac{\Lambda^2}{16 \pi^2}\, \Big( \hat{\phi}_1^\star + \hat{\phi}_1 \Big)~,
\ee 
so that the one-loop potential in terms of $\hat{\phi}_1$, $\hat{\phi}_2$ becomes
\bea 
U^{(1)} \ &=&\ M_b \frac{\Lambda^2}{16 \pi^2} \left( \hat{\phi}_1 + \hat{\phi}_1^\star \right)\: +\: m_2^2 M_c \hat{\phi}_2\: -\:\mu^2 M_c \hat{\phi}_1\: 
+\: \left( M_1^2 + \frac{g \Lambda^2}{16 \pi^2} \right)|\hat{\phi}_1|^2\: +\: m_2^2 |\hat{\phi}_2|^2 \nonumber\\
&&\qquad +\: \mu^2 \left( \hat{\phi}_2 \hat{\phi}_1^\star - \hat{\phi}_1 \hat{\phi}_2^\star \right) \:+\: \left( \frac{M_a^2}{2} 
- \left( M_b^2 + \frac{gM_a^2}{2}  \right)\frac{\ln\left( \frac{\Lambda}{m} \right)}{16 \pi^2} \right) 
\left( \hat{\phi}_1^2 + \left(\hat{\phi}_1^\star\right)^2 \right) \nonumber\\ 
&&\qquad+\: \frac{M_b}{2} \left( 1 - \frac{5g\ln \left( \frac{\Lambda}{m} \right)}{16 \pi^2} \right)|\hat{\phi}_1|^2 \left( \hat{\phi}_1^\star + \hat{\phi}_1 \right) 
\:+\: \frac{g}{4}\, \left(1 - \frac{5g \ln \left( \frac{\Lambda}{m} \right)}{16 \pi^2} \right)|\hat{\phi}_1|^4~.
\eea
To show the existence of the Goldstone mode to one-loop order, we should express the fields in terms of fluctuations around the new shifted vacuum. From this, we can find the one-loop-corrected vevs
\be 
\begin{pmatrix}
v_1^{(1)} \\ v_2^{(1)}
\end{pmatrix} \ =\ \left( 1 - \frac{g}{2M_a^2}\,\frac{\Lambda^2}{16 \pi^2} \right) \begin{pmatrix}
v_1 \\ v_2
\end{pmatrix}~.
\ee
Expressing the one-loop potential in terms of the fields fluctuating around this minimum 
\be
\phi_1 \ =\  v^{(1)}_1\: +\: \hat{\phi}_1^{(1)} \qquad\mbox{and}\qquad 
\phi_2 \ =\ v^{(1)}_2\: +\: \hat{\phi}_2^{(1)}
\ee
gives equations of motion of the form
\be
\Box \begin{pmatrix}
\hat{\phi}_1^{(1)} \\ \left( \hat{\phi}_1^{(1)} \right)^\star \\ \hat{\phi}_2^{(1)} \\ \left( \hat{\phi}_2^{(1)} \right)^\star
\end{pmatrix}\ =\ \begin{pmatrix}
M_1^2 - \frac{g \Lambda^2}{16 \pi^2} & M_a^2 - \frac{g \Lambda^2}{16 \pi^2} & \mu^2 & 0 \\ 
M_a^2 - \frac{g \Lambda^2}{16 \pi^2} & M_1^2 
- \frac{g \Lambda^2}{16 \pi^2} & 0 & \mu^2 \\ -\mu^2 & 0 & m_2^2 & 0 \\ 
0 & -\mu^2 & 0 & m_2^2
\end{pmatrix} \begin{pmatrix}
\hat{\phi}_1^{(1)} \\ \left( \hat{\phi}_1^{(1)} \right)^\star \\ \hat{\phi}_2^{(1)} \\ \left( \hat{\phi}_2^{(1)} \right)^\star
\end{pmatrix} \:+\: \cdots~.
\ee
The mass matrix again has determinant zero, showing that we still have a Goldstone mode at one-loop order, which is given by 
\be 
G_1^{(1)} \ =\  \sqrt{\frac{2 m_2^4}{m_2^4+\mu^4}}\,\bigg( \text{Im}\,\hat{\phi}_1^{(1)}\: -\: \frac{\mu^2}{m_2^2}\,\text{Im}\,\hat{\phi}_2^{(1)}\bigg)~.
\ee
We see that the one-loop Goldstone mode is related to the Goldstone mode at tree level; the one-loop mode is obtained from the 
tree-level one simply by making the replacement $\hat{\phi}_1, \hat{\phi}_2 \rightarrow \hat{\phi}_1^{(1)}, \hat{\phi}_2^{(1)}$.

\section{Summary and open questions}
\label{sec:conx}

The nature of spontaneous symmetry breaking is a fascinating and deep issue in quantum field theory. In conventional Hermitian QFT,
it is well understood how the spontaneous breaking of a global symmetry is accompanied by the appearance of a massless
scalar Goldstone boson. The counterpart of the Goldstone theorem in $\mathcal{PT}$-symmetric QFT presented certain puzzles
and has not been known until now. The central issue was that, although $\mathcal{PT}$-symmetric theories may contain conserved
currents, there are no corresponding symmetries of the Lagrangian: Noether's theorem does not apply~\cite{Alexandre:2015oha} in the familiar sense. One could then wonder
whether the existence of a conserved current would be sufficient to guarantee the appearance of a Goldstone boson, or not?

We have shown in this paper that the answer is yes: current conservation still guarantees the existence of a massless boson.
We have demonstrated this formally and also at the tree and one-loop levels in a simple $\mathcal{PT}$-symmetric QFT with
two complex scalar fields.

In order to investigate the Goldstone theorem in a $\mathcal{PT}$-symmetric theory, we studied the formulation of the
path integral and its quantisation in non-Hermitian field QFT. Since a $\mathcal{PT}$-symmetric theory possesses
a complete set of real energy eigenstates, its path integral contains saddle points about which the path
integration of quantum fluctuations is well defined, as long as one considers $\mathcal{PT}$-conjugate pairs of fields 
instead of Hermitian-conjugate pairs.

The analysis in this paper can be regarded as the first step in an exploration of whether there exists a consistent
$\mathcal{PT}$-symmetric generalisation of the Standard Model and other gauge theories. In this connection, the absence of a
generalisation to non-Hermitian theories of Noether's theorem is a key issue. We emphasise again that, in these theories,
the existence of a conserved current does not imply the existence of a corresponding symmetry. How do gauge theories
react to this situation and, in particular, do they possess a `Higgs phase'? We plan to address these issues in
future work.

\section*{Acknowledgements}

The work of JA, JE and DS was supported by the United Kingdom Science and Technology Facilities Council (STFC) Grant No.~ST/P000258/1, and that of JE also
by the Estonian Research Council via a Mobilitas Pluss grant. The work of PM was supported by STFC Grant No.~ST/L000393/1 and a Leverhulme Trust Research Leadership Award.

\appendix

\section{Running Couplings}\label{AppRunCop}

The full bare potential is
\bea\label{fullpot}
U^{(0)} \ &=&\ m_1^2|\phi_1|^2\:+\:m_2^2|\phi_2|^2\:+\:\mu^2(\phi_1^\star\phi_2-\phi_2^\star\phi_1)\nonumber\\
&&\qquad +\:\frac{g_1}{4}\,|\phi_1|^4 \:+\: \frac{g_2}{4}\, |\phi_2|^4 
\:+\: \lambda |\phi_1 \phi_2|^2\:+\: \frac{\alpha}{4}\, \Big(\left( \phi_1^\star \phi_2 \right)^2 + \left( \phi_2^\star \phi_1 \right)^2 \Big) \nonumber\\
&&\qquad+\: \frac{1}{2}\,\Big(  \beta_1 |\phi_1|^2 + \beta_2 |\phi_2|^2 \Big)\Big( \phi_1^\star \phi_2 - \phi_2^\star \phi_1 \Big)~,
\eea
and the one-loop 1PI potential is given by
\be 
U^{(1)} \ =\  U^{(0)}\: +\: \frac{1}{2V^{(4)}}\,\text{STr}\,\ln S^{(2)}_E~,
\ee
where 
\be 
S^{(2)}_E \ =\ 
\begin{pmatrix}
p^2 + U_{11^\star}^{(0)} & U_{11}^{(0)} & U_{1 2^\star}^{(0)} & U_{12}^{(0)} \\ U_{1^\star 1^\star}^{(0)} & p^2 + U_{1^\star 1}^{(0)} 
& U_{1^\star 2^\star}^{(0)} & U_{1^\star 2}^{(0)} \\ U_{2 1^\star}^{(0)} & U_{21}^{(0)} & p^2 + U_{22^\star}^{(0)} & U_{22}^{(0)} \\ 
U^{(0)}_{2^\star 1^\star} & U_{2^\star 1}^{(0)} & U_{2^\star 2^\star}^{(0)} & p^2 + U^{(0)}_{2^\star 2} 
\end{pmatrix}~,
\ee
with
\be 
U_{i^{(\star)} j^{[\star]}}\ =\ \frac{\delta^2 U}{\delta \phi_i^{(\star)} \delta \phi_j^{[\star]}}~.
\ee
We have then that
\bea
&&\frac{1}{p^{8}}\, \det S^{(2)}_E 
\ =\ 1\: +\: \frac{2}{p^{2}} \left(U^{(0)}_{11^{\star}} + U^{(0)}_{22^{\star}}\right) \nonumber\\
&&\qquad +\: \frac{1}{p^{4}}\left((U^{(0)}_{11^{\star}})^{2} + (U^{(0)}_{22^{\star}})^2+4U_{11^{\star}}^{(0)}U_{22^{\star}}^{(0)}- U^{(0)}_{11}U^{(0)}_{1^{\star}1^{\star}} 
 -U^{(0)}_{22}U^{(0)}_{2^{\star}2^{\star}}- 2U^{(0)}_{12}U^{(0)}_{1^{\star}2^{\star}} - 2U^{(0)}_{12^{\star}}U^{(0)}_{1^{\star}2}\right)\nonumber\\&&\qquad
 +\:\mathcal{O}\left(\frac{1}{p^{6}}\right)~,
\eea
such that, up to finite terms,
\bea \label{eq:evol}
&&\frac{1}{2 V^{(4)}}\,\text{STr}\,\ln S^{(2)}_E 
\ =\ \frac{1}{8 \pi^{2}} \int {\rm d}p\;p \left(U^{(0)}_{11^{\star}} + U^{(0)}_{22^{\star}}\right) \nonumber\\
&&\ -\int \frac{{\rm d}p}{16 \pi^{2}p} \,\Big( (U^{(0)}_{11^{\star}})^{2} + (U^{(0)}_{22^{\star}})^2+ U^{(0)}_{11}U^{(0)}_{1^{\star}1^{\star}} 
 +U^{(0)}_{22}U^{(0)}_{2^{\star}2^{\star}}+ 2U^{(0)}_{12}U^{(0)}_{1^{\star}2^{\star}} + 2U^{(0)}_{12^{\star}}U^{(0)}_{1^{\star}2} \big)~,\eea
and substituting the potential (\ref{fullpot}) into this expression gives the one-loop corrections (\ref{oneloopcorrections}).

\end{document}